Salt-rejecting continuous passive solar thermal desalination via convective flow and thin-film condensation

Patrick I. Babb*, S. Farzad Ahmadi*, Forrest Brent, Ruby Gans, Mabel Aceves Lopez, Jiuxu Song, Qixian Wang, Brandon Zou, Xiangying Zuo, Amanda Strom, Jaya Nolt, Tyler Susko, Kirk Fields, Yangying Zhu

* These authors contributed equally to this work
Corresponding author: Yangying Zhu, yangying@ucsb.edu

**Abstract**

Passive solar desalination is an emerging low-cost technology for fresh water production. State of the art desalinators typically evaporate water using wicking structures to achieve high solar-to-vapor efficiency by minimizing heat loss. However, wicking structures cannot reject salt continuously which limits the operating duration of the desalinators to several hours before the devices are turned off to reject salt. While significant research has focused on developing efficient evaporators to achieve high solar-to-vapor efficiency, inefficient condensers have become the bottleneck for the overall solar-to-water efficiency. To overcome these challenges, we designed a passive inverted single stage solar membrane desalinator that achieves continuous desalination and salt rejection with a novel architecture. By flowing salt water on a radiative absorbing, porous, hydrophobic evaporator membrane using gravity, salt continuously diffuses away from the membrane while allowing heated water vapor to transport to and condense on a cooler microporous membrane below. In addition, our design utilizes thin-film condensation on a microporous membrane which offers ample three-phase contact region to enhance condensation phase change heat transfer. By condensing within the microporous membrane instead of forming droplets, we reduce the gap distance between the condenser and evaporator membranes, which effectively reduces the vapor transport resistance. We experimentally demonstrated a record-high continuous desalination and salt rejection test duration of 7 days under one-sun illumination, while prior state of the art work demonstrated 12 hours of continuous desalination due to salt accumulation on the wicking evaporator. Despite an increased convection heat loss necessary for salt rejection on the evaporator, our desalinator still achieved a water-collection rate of 0.487 kg m$^{-2}$ h$^{-1}$, which corresponds to a 32.2% solar to water efficiency. This work signifies an improvement in the robustness of current state of the art desalinators and presents a new architecture to further optimize passive solar desalinators.

**Introduction**
Water and energy are among the greatest challenges of the 21st century, yet current industrial-scale desalination technologies including reverse osmosis and distillation are still energy-intensive.[1] Solar thermal desalination is an attractive passive technology where sunlight is utilized as heat to vaporize water[2] which can then be condensed to pure liquid water. These devices are potentially cost-effective and portable, making them suitable for developing countries where water scarcity is most severe.[3] So far, most previous works have focused on engineering solar-absorbing porous materials to localize solar heat to the liquid-vapor interface.[4–11] Highly-efficient evaporation can be achieved by avoiding heating the bulk of water and minimizing conduction, convection, and radiation heat loss. Solar-to-vapor efficiency, usually defined as the total latent energy required to produce a certain mass of vapor over the input sunlight,[12] exceeding 90% has been demonstrated under concentrated and one sun conditions.[10]

Despite the advancements in evaporator development, only a few recent works have incorporated condensation, which is a necessary step to convert water vapor back to liquid water, into their desalination systems.[13–18] Most of these full systems have an inverted design, where the condenser is below the evaporator to avoid blocking the sunlight. To reduce the thickness of liquid in the evaporator, several inverted systems employed a thin wicking layer to feed salt water using capillarity to the solar absorber for evaporation. The minimum amount of water in the wicking layer significantly reduced heat loss. As such, solar-to-water efficiency as high as 70% has been demonstrated under one sun for single-stage devices, and even higher solar-to-water efficiencies are achieved with multi-stage devices. Solar-to-water efficiency is evaluated using the mass of collected liquid water as opposed to the mass of the evaporated water vapor in the solar-to-vapor efficiency. However, the wicking mechanism inherently lacks a salt rejecting mechanism, therefore, the high efficiencies are mainly demonstrated with pure water. With salt water, salt crystals were observed on the evaporator as soon as two hours after being exposed to sunlight.[15] Several recent salt-rejecting designs utilized a contactless absorber[8] and salt diffusion[8,15,19], yet most of these systems only have the evaporator with no condenser. A full system with a salt rejection mechanism that allows continuous desalination without efficiency decay is highly desirable.

Furthermore, most previous work incorporated dropwise condensation on hydrophobic surfaces. While dropwise condensation has a higher heat transfer coefficient than the conventional film-wise condensation, it relies on a hydrophobic coating which has been known to suffer from durability issues. Dropwise condensation also relies on gravity to remove the condensed drops and therefore cannot be positioned horizontally. In fact, several previous studies reported droplets stuck to the condenser surface which reduced the amount of collected liquid.[14,15] In addition, efficient vapor transport from the evaporator to the condenser is also critical to ensure a high solar-to-water efficiency. This requires a minimal gap distance between the evaporator and the condenser surface, assuming that there is sufficient thermal isolation between the evaporator and the condenser. This is difficult to achieve with condenser surfaces that are not parallel to the evaporator.

In this work, we designed a full system with the following key considerations as shown in Figure 1. On the evaporator side, salt water flows from a reservoir into the evaporator driven by gravity. A thin liquid layer is maintained on the top of a hydrophobic, solar-absorbing membrane. Sunlight is absorbed as heat on the membrane, which evaporates water. Evaporation increases the local concentration of salt, which diffuses back to the bulk flow and the higher-salinity water is discharged from the evaporator. This discharge mechanism allows the amount of salt within the evaporator to reach equilibrium over time, as opposed to continuously increasing in a wick structure. This allows continuous desalination without salt precipitation. However, this design will inevitably increase heat loss due to natural convection within the water layer and the sensible heat associated with the discharged water. To reduce heat loss, we recycle the heat of the discharged water to preheat the incoming water by designing a counter-flow heat exchanger.

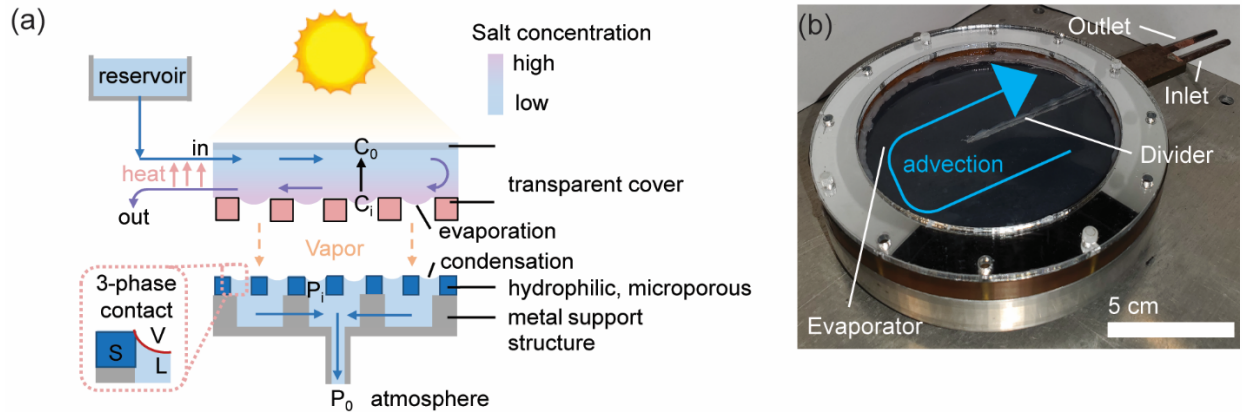

Figure 1. Design of the salt-rejecting desalination concept. (a) Cross section drawing of the desalination design and transport mechanisms. (b) Optical image of the desalinator. Blue arrow indicates the path of fluid flow through the device over the evaporator. The divider, composed of PETG and the inlet/outlet of the device are also shown. The warm water leaving the outlet preheats the cool water entering the inlet.

On the condenser side, we designed a microporous membrane for highly efficient thin-film condensation as shown in Figure 1a. Different from conventional drop-wise and film-wise condensation on a solid surface where droplets condense and shed on the same side of the surface, the vapor condenses on the front side of the membrane and the condensed liquid is collected from the backside. This way, the condenser membrane can theoretically be placed extremely close to the evaporator membrane, allowing minimal vapor transport resistance. Analogous to thin-film evaporation from micro/nanostructures, which achieves the highest phase change heat transfer coefficient, thin-film condensation in microporous membrane allows abundant three-phase contact region (Figure 1a) and can therefore potentially achieve a higher condensation heat transfer coefficient compared to drop-wise condensation, if designed properly. Furthermore, the condenser membrane is hydrophilic and capillary pressure was utilized to lower the local liquid pressure within the pores. This promotes the condensed liquid to spontaneously enter the backside of the membrane, which prevents flooding of liquid on top of the membrane. This low pressure is created simply via a hydrostatic pressure between the membrane and the outlet of the condensed liquid, which is at atmospheric pressure.

The passive, single-stage device achieved continuous sea water desalination under one sun for 1 week (168 hours) without any efficiency decay signifying, insignificant salt accumulation on the evaporator. To the best of our knowledge, this work demonstrates the longest passive continuous salt water to fresh water distillate production to date compared with state-of-the-art solar thermal desalinators. [9,13–15,18,20] The maximum solar to water efficiency measured on sea water is 32.2%. The majority of heat loss occurred in the evaporator due to natural and forced convection which are necessary for salt rejection.

**Results and Discussion**

The desalinator consists of three primary components: the reservoir, the evaporator and the condenser, as shown in Figure 1a. Liquid flows from the reservoir into the evaporator driven by gravity. An ethylene tetrafluoroethylene (ETFE) transparent cover transmits sunlight to a microporous solar-absorbing and hydrophobic evaporator membrane. Salt water is sandwiched between the ETFE transparent cover and the evaporator membrane. The evaporator membrane was fabricated by coating carbon black nanoparticles onto a commercially available gas diffusion layer (GDL) using polydimethylsiloxane (PDMS) with curing agent as the binder (Figures 2a-b). The GDL is a fibrous, Teflon treated carbon paper with a porosity of 78%.[21] Two copper tubes serve as the water inlet and outlet of the evaporator in order to maintain a salt concentration in the evaporator chamber below saturation. The copper tubes are thermally connected so that exiting warm water preheats the incoming cool water to recover thermal energy. Above the evaporator there is a polyethylene terephthalate glycol (PETG) divider to separate the inlet from the outlet such that liquid will flow across the entire evaporator membrane before being discharged (Figure 1b).

The evaporator and condenser are separated by a thin air gap (3 mm, using a rubber O-ring) which allows efficient vapor transport from the evaporator to the condenser while still maintaining a temperature difference. A thinner gap distance reduces the vapor transport resistance which is modeled using the Fick's law (see supplemental information for details) but may increase the heat conduction via the O-ring from the evaporator to the condenser. In addition, as the temperature of the water above the evaporator increases, the solubility of air in the water decreases which can cause air bubbles to form on the evaporator membrane. In order to mitigate this problem, a small opening is introduced on the O-ring to allow air to vent. The condenser membrane is a hydrophilic microporous paper (pore size approximately 100-580 µm) made of 55% cellulose and 45% polyester (Figure 2c, Texwipe TX609)[22] which provides ample three-phase contact area for effective condensation. To reject heat, the microporous paper is supported by thermally conductive nickel foam attached to aluminum, where heat is dissipated to the environment through natural air convection. The larger pore size of the nickel foam (Figure 2d) allows the condensed fresh water to flow laterally with a low viscous resistance to an outlet positioned at the center of the device (Figure 1a and SI Figure S1).

We characterized the optical property of the membranes and films through Ultraviolet-Visible-Near-Infrared (UV-Vis-NIR) spectrophotometry. Figure 2e shows that the evaporator membrane has an absorptivity, $\alpha_{evap}$ greater than 0.9 for the majority of the solar spectrum between 200 nm and 2.6 µm and an absorptivity greater than 0.85 for the solar spectrum between 200 nm and 2.6 µm. This result shows that the evaporator membrane is absorbing the vast majority of incoming solar light and heating since about 99% of solar radiation is contained in the region between 300 nm to 3 µm.[23] Figure 2f illustrates that ETFE, the transparent cover for the device, has a transmittance $\tau_{ETFE}$ of greater than 0.9 for the majority of the spectrum between 200 nm and 2.6 µm, which allows for high solar transmission to the membrane. Since the layer of water sandwiched between the evaporator and ETFE is thin (~ 1 mm), we estimate that less than 80% of the incoming solar energy ($\tau_{ETFE} \times \alpha_{evap}$) is used to heat the evaporator membrane. The thickness of water increases to 5 mm at the inlet and outlet to allow water from the copper tubes to enter the desalinator freely. The divider has a lower transmittance than ETFE that declines after 1.5 µm however the divider is only exposed to a negligibly small portion of the evaporator membrane.

Before testing the full system, the liquid to vapor efficiency of the evaporator was first characterized under a solar simulator (Newport ORIEL SOL3A) with continuous solar flux of 1000 W m$^{-2}$ (one sun) for 10 hours. The device has only the evaporator which evaporates water into ambient air (inset, Figure 2g). The diameter of the exposed evaporator membrane is 9 cm which corresponds to an area of 6.362×10$^{-3}$ m$^2$. We used salt water obtained from Goleta Beach in California. The salinity was 2.90 wt%, measured using an ORAPXI digital salinity tester (PS-A200). In order to estimate an upper limit of the solar to vapor efficiency corresponding to a minimal liquid flow rate, we used a syringe pump to supply salt water to the evaporator and gradually reduced the flow rate until the flow rate of the discharged water was negligibly low. At this condition, the inlet flow rate was 4.9 mL h$^{-1}$ and the discharged flow rate was measured to be 0.18 mL h$^{-1}$. The mass of the evaporated liquid over time shown in Figure 2g was calculated by subtracting the mass of the discharged water measured using a mass balance from the mass of the supplied liquid. Although this discharge flow rate may be too low to avoid salt precipitation on the evaporator membrane over an extended period of time, the salinity of sea water on the evaporator after the 10 hour test concluded was 3.32 wt%, well below the saturation concentration of 26.89 wt%. This saturation concentration is evaluated at 50 °C, which was roughly the evaporator temperature. [24] No salt precipitation was observed on the evaporator membrane immediately after the experiment.

As seen in Figure 2g, the evaporator membrane alone has a vapor production rate of 4.72 g h$^{-1}$ or 0.742 kg m$^{-2}$ h$^{-1}$ corresponding to a solar to vapor efficiency of 49.1 %. The definition of efficiency is $\eta = \frac{\dot{m} h_{fg}}{q''}$ where $\dot{m}$ is the mass flow rate, $h_{fg}$ = 2381.9 kJ kg$^{-1}$ is the latent heat of vaporization at 50 °C. We chose the latent heat at 50 °C, because this is the approximate evaporator temperature we observed in subsequent experiments using the same evaporator membrane and ETFE transparent film. The incoming solar flux is $q''$ = 1000 W m$^{-2}$ (one sun). Only latent heat, not sensible heat is considered in order to provide a conservative solar to vapor efficiency result. The majority of heat is lost through natural and forced convection in the thin liquid layer on top of the membrane. Thus, the evaporator efficiency could be improved by optimizing the liquid layer thickness to reduce convective heat loss while still allowing salt diffusion, as well as improving thermal insulation of the top transparent cover. Many recent works on the evaporator without the condenser reported solar to vapor efficiency above 85%[6,25–33]. The high efficiency was typically achieved by positioning a membrane on top of liquid, so the natural convection in the liquid is suppressed, effectively localizing heat to the membrane.

However, such design if made into a full system would require a condenser on top of the evaporator, which is impractical as the condenser will block sunlight.

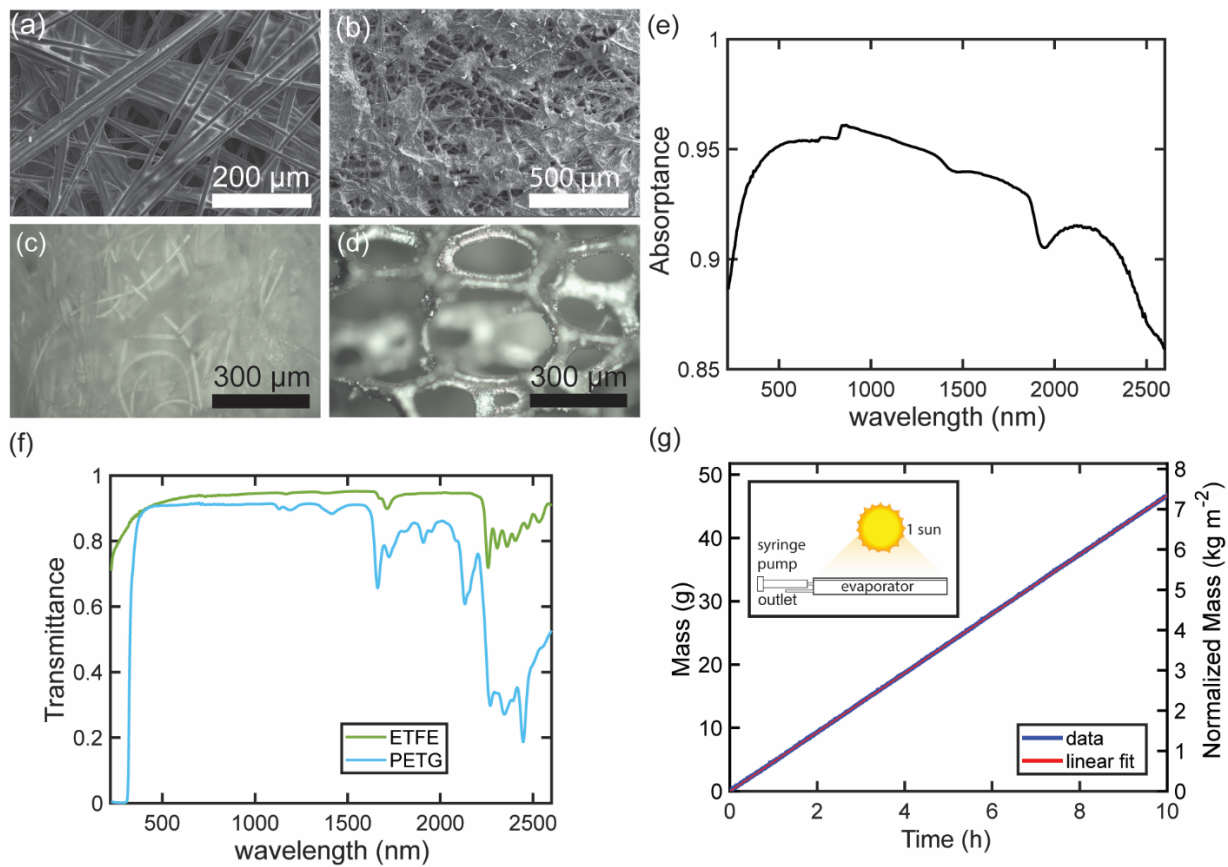

Figure 2. Characterization of optical properties and vapor production of the evaporator. (a) Scanning electron microscopy (SEM) image of the GDL substrate alone. (b) SEM image of the evaporator membrane composed of the GDL substrate coated with PDMS and carbon black nanoparticles. (c) Optical image of the microporous paper condenser membrane. (d) Optical image of the nickel foam. (e) Absorptivity of the evaporator membrane using the UV-Vis-NIR spectrophotometer. (f) Transmittance spectra of the ETFE and PETG films used as a transparent cover and divider respectively. (g) Vapor production from the evaporator membrane is 4.72 g h$^{-1}$ corresponding to 0.742 kg m$^{-2}$ h$^{-1}$. The inset figure illustrates the evaporator used without the condenser to measure the vapor flux.

We then constructed a full system containing the evaporator and the condenser and conducted a prolonged desalination test lasting 7 days using salt water from Goleta Beach in California. Figure S2 illustrates the test setup. The tests were performed under one sun condition using the solar simulator. The measured salinities of salt water ranged from 2.65 wt% to 2.95 wt% during the duration of the test. As a benchmark test, we used a syringe pump to provide a well-controlled inlet flow rate of 5 mL h$^{-1}$ to rule out the influence of unsteady flow rate on the solar to water efficiency. As we show in a later section, fluid can be supplied passively using a gravity feeding bag. The flow rate of 5 mL h$^{-1}$ was determined to be sufficient in avoiding salt precipitation as the concentration of salt water from the outlet was 5.60 wt% after 7 days, which was lower than the saturation salt concentration of 26.889 wt%. More information on the inlet flow rate is in the supplemental information Note S2 and Figure S3.

Figure 3a illustrates the distillate production as a function of time which was measured by a mass balance. The rate of fresh water collection does not decay even after 7 days (168 hours) under one sun illumination, which indicates that the pore sizes of the evaporator membrane possibly remained unchanged and we further hypothesize that no salt crystals were formed. Figure 3b demonstrates temperatures of the evaporator membrane, condenser membrane as well as the ambient temperature of the laboratory during the experiment. The temperature difference between the evaporator and the condenser was less than 10 °C. Based on the temperature measurement, we used Fick's law to determine the theoretical distillate production and distillate rate. Lennard Jones potentials were used to determine the diffusivity as a function of temperature which is described in detail in the supplemental information (Note S1). The mass flux from the evaporator membrane to the condenser membrane can therefore be derived due to a vapor pressure gradient. Figure 3c compares distillate production obtained from the laboratory results to the theoretical distillate production based on the temperatures of the evaporator and condenser membrane. The maximum theoretical distillate production after 7 days is 894.7 grams (140.6 kg $m^{-2}$) whereas the maximum experimentally determined distillate production after 7 days is 531.1 grams (83.5 kg $m^{-2}$). Figure 3d shows the experimentally and theoretically determined distillate production rates over the duration of the test. The distillate rates are calculated every 30 minutes based off distillate production data. The experimentally determined distillate rate is 3.1 g $h^{-1}$ which corresponds to 0.487 kg $m^{-2}$ $h^{-1}$ and a solar to water efficiency of 32.2%. The measured salinity of the distillate was 0.00 wt%. The theoretically determined average distillate rate is 5.24 g $h^{-1}$ which corresponds to 0.824 kg $m^{-2}$ $h^{-1}$ and a solar to water efficiency of 54.5%. The discrepancy between the experimental and theoretical values could originate from the uncertainties in the measured temperature and the distance between the evaporator and the condenser, leading to less than accurate theoretical prediction. In addition, the spatial heterogeneity in the device was also not captured by the single-point thermocouple measurements, and there exists minor leakage of water vapor in the experimental setup due to the degassing design.

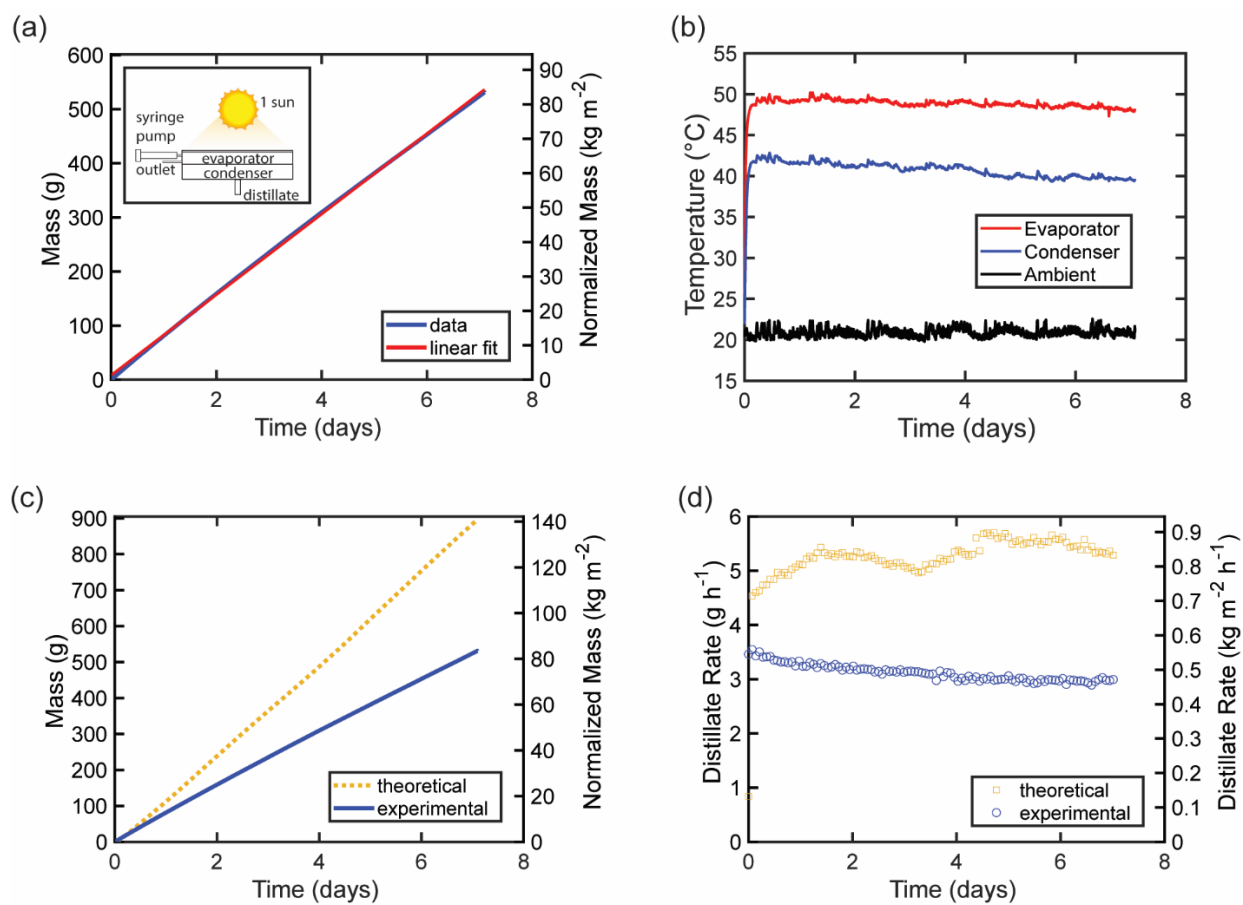

Figure 3. Full system performance over 7 days using 2.65 wt% to 2.95 wt% salinity seawater and solar flux of 1000 W m$^{-2}$. Measured distillate salinity is 0.00 wt%. (a) Continuous distillate production over 7 days with a salt water inlet flow rate of 5 mL h$^{-1}$. The schematic of the laboratory test setup is inset (b) Evaporator, condenser and ambient temperatures over 7 days. (c) Theoretical distillate production based on Fick's Law and experimental distillate production over 7 days. (d) Theoretical distillate production rate based on Fick's Law and experimental distillate production rate.

In order to move on to fully passive tests, a gravity feeding bag was substituted in place of the syringe pump in order to achieve passive gravity driven flow, as illustrated in Figure 4a. We determined the flow rate from the gravity feeding bag by determining the distillate rate and outlet flow rate from the distillate mass balance and outlet mass balance. The flow rates are calculated every 30 minutes using distillate production data. The exposed evaporator area is 7.585 ×10$^{-3}$ m$^2$ which is slightly larger than the exposed evaporator area from the 7-day test. As seen in Figure 4b, we achieved a distillate production rate of 2.87 g h$^{-1}$ which corresponds to 0.378 kg m$^{-2}$ h$^{-1}$ over 25 hours in one sun using the passive method and a 25.6 % solar to water efficiency. This efficiency is slightly lower than the 7 day test because of the higher flow rate which leads to a higher convective heat loss. The temperatures of the evaporator and the condenser membranes are shown in Figure 4c. The temperature on the outside surface of the condenser is approximately 15 °C above the ambient. This indicates that the air natural convection thermal resistance on the external surface of the condenser was the dominant resistance, which is higher than the thermal resistances of vapor transport resistance and condensation. Further reducing the thermal resistance from the condenser surface to the ambient could help lowering the temperatures of the condenser and evaporator membranes and thus reduce heat loss. As shown in Figure 4d, the gravity feeding bag flow decreased in 3 cycles during the course of 25 hours due to a decrease in the available hydrostatic pressure as salt water flows from the gravity feeding bag to the desalinator. The average flow rate over the duration of the test was 23.5 g h$^{-1}$. The flow rate from the gravity bag was increased by manually opening a needle valve attached to the gravity bag at hour 9 and hour 18 of the experiment to prevent dryout of liquid in the evaporator. This led to temperature drops at hour 9 and hour 18 in the evaporator and condenser membranes (Figure 4c). The solar to water efficiency of the passive desalination test was only 6.6% less than the controlled syringe pump tests seen in figure 2, indicating the viability of this passive method.

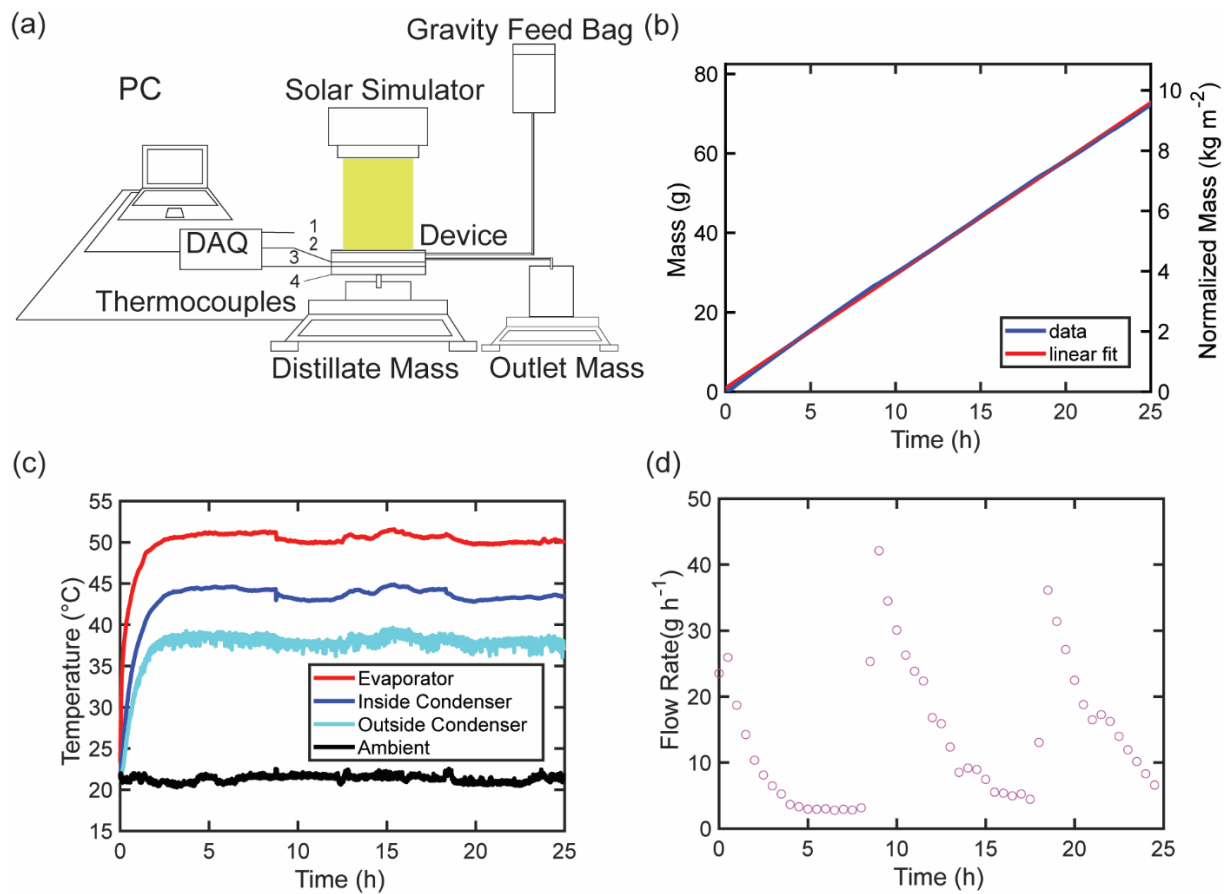

Figure 4. Full system performance using a passive gravity feed bag for flow. (a) Schematic of the passive desalination setup. (a) Linear distillate production over 25 hours. Inlet salt water salinity varies from 2.49 wt% to 2.65 wt% salinity seawater and solar flux is 1000 W m$^{-2}$. Measured distillate salinity is 0.00 wt% (c) Evaporator, condenser and ambient temperatures over 25 hours. Thermocouple 1 measures the ambient temperature. Thermocouple 2 measures the temperature of the evaporator. Thermocouple 3 measures the temperature on the surface of the condenser directly beneath the evaporator. Thermocouple 4 measures the temperature on the outside surface of the condenser. (d) Flow rates entering desalinator over 25 hours.

**Conclusions**

This 7-day desalination result is significant because it is the longest continuous solar desalination with passive salt rejection to date. Other works such as one by Wang et al. report continuous solar desalination with no salt accumulation for 12 hours when using 7 wt% NaCl.[13] One work by *Xu et* al reports salt accumulation and rejection over 18.5 hours with illumination for 3 hours using 3.5 wt% NaCl.[14] In addition, Ni et al. reports continuous solar vapor production from 3.5 wt% NaCl for 2000 seconds[15] (~0.55 hours). These works reported these results under laboratory conditions and indicate that the result from our work is a contribution to the field.

Efficiency of our device can be improved by reducing convective heat loss from the evaporator membrane to the salt water layer by further reducing the water layer thickness and flow rate, and by reducing the convective heat loss from the transparent cover to the ambient environment. Efficiency can also be improved by reducing conduction from the warm evaporator to the cold condenser, while maintaining a low vapor transport resistance. In addition, the target flow rate of 5 mL $h^{-1}$ may be further reduced. At 5 mL $h^{-1}$, the equilibrium concentration of the salt water in the desalinator was 5.60 wt% which is below saturation (26.889 wt%). At a reduced flow rate, a higher equilibrium evaporator temperature can potentially be established leading to higher vapor and distillate production. On the condenser side, enhancing heat rejection from the aluminum surface to the ambient environment may also help to lower the condenser temperature. This can be achieved with increased aluminum surface area, or incorporating radiative cooling. Furthermore, we introduce a novel thin-film condensation approach, which utilizes the highly effective three-phase contact region for phase change heat transfer, as well as a low vapor transport resistance by reducing the distance between the evaporation and the condensation surfaces. By demonstrating 7 days of continuous desalination, we have achieved the longest continuous desalination among the reported solar thermal desalination full systems, with salt rejection under 1 sun. This work presents a new architecture that addresses a longstanding problem of salt accumulation in current state of the art solar thermal desalination devices, and highlights the importance of optimizing vapor transport and condensation in the overall efficiency of passive solar desalinators.

## Methods
**Fabrication of the evaporator membrane**

To enhance the optical absorptivity of the evaporator membrane, we followed a method by reported by Liu et al.[34] The evaporator membrane is composed of two processes, membrane preparation and preparation of the coating solution. First the membrane is prepared. 0.8 g of carbon powder (Sid Richardson Carbon & Energy Co.) is dispensed into a 250 mL beaker containing 160 mL of water. 3 mL of acetic acid is added to make the carbon powder easier to attach to fibers. The mixed solution is blended using an ultrasonic cleaner (Branson Ultrasonics BransonicTM B200) for 5 minutes. The 110 mm diameter porous carbon paper (Toray Paper 060 TGP-H-060) is added to the mixed solution to vibrate for 3 minutes in order for the carbon powders to dye the paper uniformly. We dried the carbon immersed GDL at 80 °C on a hot plate for 2 hours. We repeated the membrane preparation process four times to realize the desired dark color.[34]

According to Liu et al., the coating solution is prepared by dispersing 1 g of PDMS containing 10% curing agent and 1 g of carbon black nanoparticles into 30 mL of hexane. The coating solution is mixed by a magnetic stir bar for 10 minutes. We immersed the membrane into the coating solution. Then, we cured the membrane on a hotplate at 80 °C for 2 hours.[34]

**Experimental setup and procedure**

The grade AAA solar simulator by Newport ORIEL SOL3A illuminates the desalinator with a power flux of 1000 W/m². Saltwater is supplied to the desalinator device using a syringe pump (Cole Palmer, catalog no. 78-0100C) and a (Harvard Apparatus Pump 33DDS, Item No. 70-3333) with a controlled flow rate of 5 mL h$^{-1}$. The syringe pump drives flow actively into the desalinator inlet. To dispel salt water from the device, the inlet flow rate is greater than the distillate rate. The inlet flow is tuned to optimize efficiency and salt water rejection. The salt concentration within the desalinator does not exceed 26.889 wt%, the saturation concentration for salt in water. For passive tests, salt water is supplied using two gravity feeding bags (Denshine Disposable Enteral Nutrition Bag Feeding Bag, 1200mL Enteral Delivery Gravity/Pump Feeding Bag Set, Detachable Connector, (model no. ACT5435)) which provide flow with no energy consumption. The gravity feeding bag rate is controlled using a precision flow-adjustment valve (McMaster Carr, item no. 48965K28). A mass balance (OHAUS Scout, model no. SPC621) is used in conjunction with OHAUS serial port data collection (SPDC) to measure the distillate mass at a 10 second sample frequency. Thermocouples (Omega J-Type, model no. 5SRTC-TT-J-36-36) are used to measure the evaporator, condenser and ambient air temperatures.

In order to measure the vapor flux observed in Figure 2g, we supplied sea water with a salinity of 2.90 wt% to the desalinator device using a syringe pump for 10 hours. In order to estimate an upper limit of the solar to vapor efficiency corresponding to a minimal liquid flow rate, we used a syringe pump to supply salt water to the evaporator and gradually reduced the flow rate until the flow rate of the discharged water was negligibly low. At this condition, the inlet flow rate was 4.9 mL h$^{-1}$ and the discharged flow rate was measured to be 0.18 mL h$^{-1}$. The mass of the evaporated liquid over time shown in Figure 2g was calculated by subtracting the mass of the discharged water measured using a mass balance from the mass of the supplied liquid. The salinity of the sea water on the evaporator membrane was 3.32 wt% after 10 hours and no salt crystals formed on the evaporator membrane during this duration. The detailed experimental setup is shown in Figure S4.

## Author contributions

P.B. and F.A. contributed equally to this work. F.A. and Y.Z. ideated and developed the initial prototype. P.B. performed experiments, data analysis and modeling for this work. F.B., R.G., M.A.L., J.S., Q.W., B.Z. designed and fabricated a scaled-up desalinator under the supervision of K.F. and T.S.. F.B. developed a prototype model, sourced materials, and led manufacturing with contributions from B.Z. and R.G. J.S. and Q.W. manufactured a test stand. M.A.L. and R.G. performed heat transfer analysis to validate design additions. X.Z. contributed to fabrication and collection of salt water. A.S., J.N. and P.B. conducted the absorptivity measurement. P.B. conducted the transmissivity measurements and took the SEM images. All authors discussed and contributed to the writing and revision of this paper.

## Notes
The authors declare no competing financial interest.

## Acknowledgments
This work was supported by the National Science Foundation under grant number 2047727 and a seed fund from Institute for Energy Efficiency at University of California, Santa Barbara. Patrick Babb acknowledges the National Science Foundation Graduate Research Fellowship Program (GRFP), grant number 2139319. The authors thank Marty Ramirez for supervising the manufacturing of the desalinator device, and Aleks Labuda who machined a preliminary desalinator. The authors acknowledge the Materials Department at UCSB for providing the SEM equipment. The authors acknowledge the Materials Research Laboratory for providing the UV-Vis NIR equipment. The MRL Shared Experimental Facilities are supported by the MRSEC Program of the NSF under Award No. DMR 1720256; a member of the NSF-funded Materials Research Facilities Network (www.mrfn.org).

# Supplementary Information

# Salt-rejecting continuous passive solar thermal desalination via convective flow and thin-film condensation

Patrick I. Babb[1*], S. Farzad Ahmadi[1,2*], Forrest Brent[1], Ruby Gans[1], Mabel Aceves Lopez[1], Jiuxu Song[1], Qixian Wang[1], Brandon Zou[1], Xiangying Zuo[1], Amanda Strom[3], Jaya M. Nolt[3], Tyler Susko[1], Kirk Fields[1], and Yangying Zhu[1]

[1]Mechanical Engineering Department, University of California, Santa Barbara, 93106, USA

[2]Department of Physics, McDaniel College, Westminster, MD 21157, USA

[3]Materials Research Laboratory, University of California, Santa Barbara, 93106, USA

* These authors contributed equally to this work

Corresponding author: Yangying Zhu, yangying@ucsb.edu

**Supplementary Note S.1**

**Mass Transport Analysis**

To understand the theoretical maximum mass transport of water through air from the evaporator to the condenser, we use Fick's law of diffusion,

$$m" = D_{H_2O,air} \frac{c(T_{evaporator}) - c(T_{condenser})}{z}$$

Where $c(T_{evaporator})$ and $c(T_{condenser})$ are the saturated vapor concentrations at the evaporator (with temperature $T_{evaporator}$) and the condenser (with temperature $T_{condenser}$). The length, z is the distance between the evaporator and the condenser. For this model, z = 3 mm. In this model, we considered the temperature-dependent mass diffusivity $D_{AB}$ of binary species, as per Chapman-Enskog theory[1,2],

$$D_{AB} = \frac{1.8583 \times 10^{-7} T^{3/2}}{p\sigma_{AB}^2 \Omega_D} \sqrt{\frac{1}{M_A} + \frac{1}{M_B}}$$

Where the value, $1.8583 \times 10^{-7}$ is an empirical coefficient, $p$ is the ambient pressure and $\sigma_{AB}$ is the average collision diameter. The air temperature, T is computed by taking the average of evaporator and condenser temperatures. $\Omega_D$ is the collision integral for diffusivity. $M_A$ and $M_B$ are the molar mass of the two species: water and air.

We compute the diffusivity, $D_{AB}$ for the diffusion of $H_2O$ in air at 318.5 K and 1 atm. Let air be species A and $H_2O$ be species B. $\varepsilon_A$ is the potential well depth for species A and $\varepsilon_B$ is the potential well depth for species B. The potential well depth is a measure of how strongly two particles attract each other. $\varepsilon_{AB}$ is a measure of how strongly species A is attracted to species B. $k_B$ is the Boltzmann constant.

For the given conditions, using table 11.2 from Lienhard IV et. al. calculated from mass diffusion data we determine the Lennard-Jones constants.[3,4]

$$\sigma_A = 3.711 \text{ Å and } \sigma_B = 2.655 \text{ Å}$$

$$\frac{\varepsilon_A}{k_B} = 78.6 \text{ K and } \frac{\varepsilon_B}{k_B} = 363 \text{ K}$$

From these constants, we can calculate the following values for the air-water mixture.

$$\sigma_{AB} = \frac{\sigma_A + \sigma_B}{2} = 3.183 \text{ Å}$$

$$\frac{\varepsilon_{AB}}{k_B} = \sqrt{\left(\frac{\varepsilon_A}{k_B}\right)\left(\frac{\varepsilon_B}{k_B}\right)} = 169 \text{ K}$$

$$\frac{k_B T}{\varepsilon_{AB}} = 1.8856$$

Hence, $\frac{\varepsilon_{AB}}{k_B} = 168.9136 \text{ K}$ and $\Omega_D = 1.0972$ using table 11.3 from Lienhard IV et. al.[3] Then,

$$D_{AB} = \frac{1.8583 \times 10^{-7} T^{3/2}}{(1)(3.183)^2(1.0972)}\sqrt{\frac{1}{28.96} + \frac{1}{18.02}} = 2.7773 \times 10^{-5} \text{ m}^2/s$$

An empirical correlation for the water air mixture can also be used.[3]

$$D_{H_2O,air} = \frac{1.87 \times 10^{-10} T^{2.072}}{p} \text{ for } 282 \text{ K} \leq T \leq 450 \text{ K}$$

Using T = 318.5 K for this empirical correlation, we get:

$$D_{H_2O,air} = 2.873 \times 10^{-5} \text{ m}^2/s$$

The prediction is within 4% of the empirical value.

The temperature of the evaporator and condenser is used to find the corresponding vapor pressure. An equation of state, the ideal gas law is used to correlate the vapor pressure to the concentration.

$$PV = nRT$$

$$PV = \frac{m}{M}RT$$

$$\frac{PM}{RT} = \frac{m}{V} = c$$

The table includes the respective concentrations for the evaporator and the condenser.

Table 1. Properties of water at the evaporator and condenser

|  | Evaporator | Condenser |
|---|---|---|
| Temperature (K) | 322 | 315 |
| Vapor Pressure (Pa) | 12194 | 8118.5 |
| Concentration (Kg m$^{-3}$) | 0.0818 | 0.0559 |

$$m" = 2.7773 \times 10^{-5} \frac{0.0818 - 0.0559}{0.003} = 2.398 \times 10^{-4} \: kg \: m^{-2} s^{-1}$$

For an illuminated evaporator area of 0.006362 m², as in the 7-day test,

$$m" = 5.492 \: g \: hr^{-1}$$

The experimental mass flux is: $m" = 3.1 \: g \: hr^{-1}$. The experimental mass flux is ~ 44% lower than the theoretical mass flux. The discrepancy may originate from uncertainties in the temperature measurement, spatial heterogeneity as well as vapor loss.

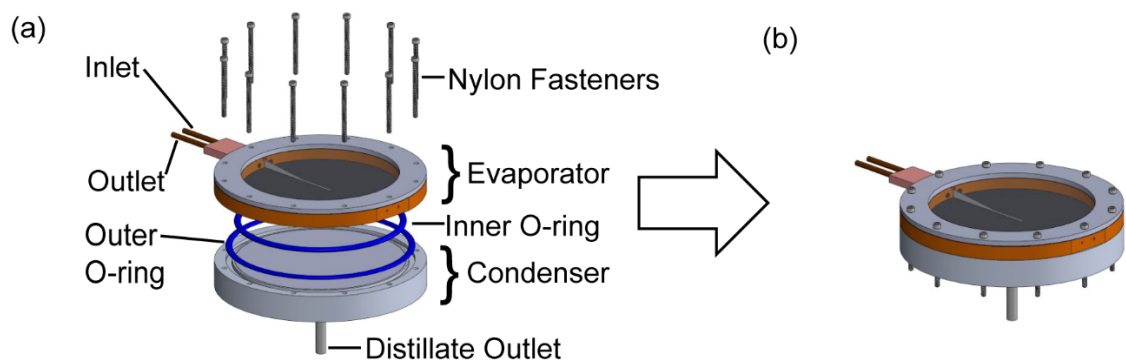

Figure S1. CAD isometric views of the desalinator. (a) Expanded view illustrating components used to assemble the desalinator. (b) Assembled desalinator used in tests.

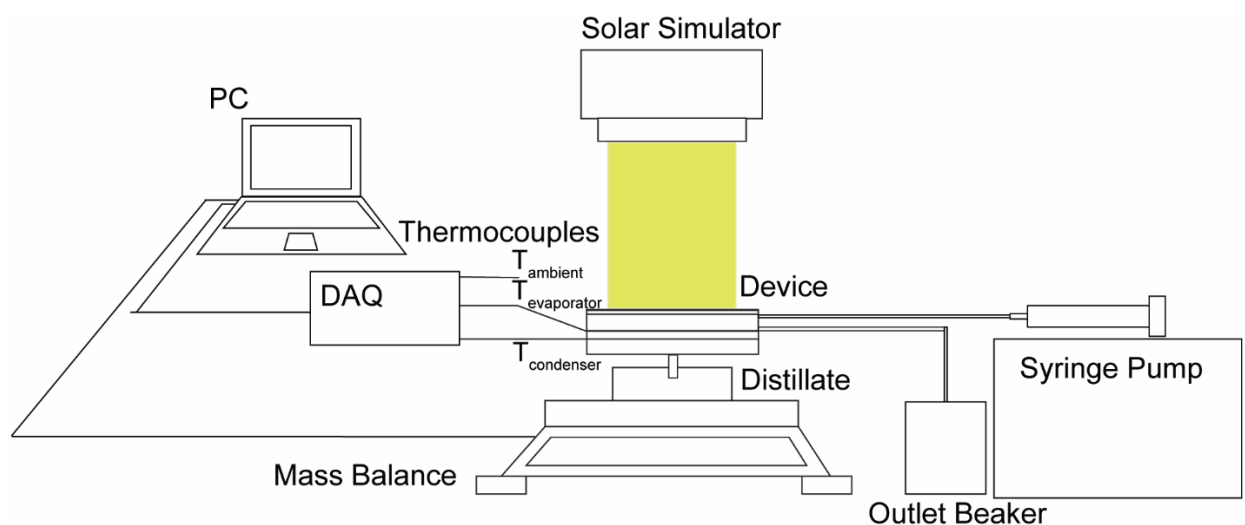

Figure S2. Schematic of the test setup used in the 7-day test. The mass data from the mass balance and temperature data from the temperature data acquisition (DAQ) system was fed into logging software on the PC. The syringe pump provided a controlled inlet flow rate of 5 ml h$^{-1}$.

**Supplementary Note S.2**

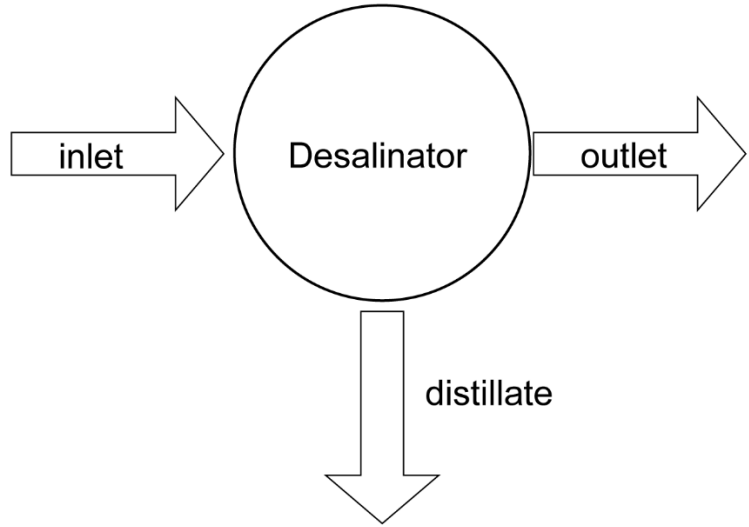

Figure S3. Schematic of the inputs and outputs for fluid flow in the desalinator. Sea water is fed into the inlet, and distilled fresh water and higher concentration salt water exits the desalinator.

Where the mass of the salt water solution entering the device is $x_1$, the mass of salt water exiting the desalinator is $x_3$, and the mass of distillate produced is $x_2$, we see that $x_1 = x_2 + x_3$.

Where the mass of the solute, salt entering the device is $m_1$, the mass of salt exiting the desalinator is $m_3$, and the mass of salt in the distillate produced is $m_2 = 0$, we see that $m_1 = m_3$.

Weight percent is defined as:

$$\frac{mass\ of\ solute}{mass\ of\ solution} \times 100.$$

Therefore, the weight percent of salt water entering the device at the inlet is:

$$\frac{m_1}{x_1} \times 100$$

The estimated weight percent of salt water exiting the device at the outlet is:

$$\frac{m_3}{x_3} \times 100 = \frac{m_3}{x_1 - x_2} \times 100$$

We define salt concentration of the salt water at the inlet as:

$$c_1 = \frac{m_1}{x_1}$$

We define the salt concentration of water at the outlet as:

$$c_3 = \frac{m_3}{x_3}$$

Thus, to estimate the expected concentration of the salt water over 7 days (168 hours) at the outlet, given an inlet flow rate of 5 g h⁻¹, an inlet salinity of 2.8 wt% and an observed initial distillate production rate of approximately 3 g h⁻¹ we make the following calculation:

$$c_3 = \frac{c_1 x_1}{x_1 - x_2} = \frac{(.028)(5\ g\ h^{-1})(168\ h)}{((5\ g\ h^{-1})(168\ h) - (3\ g\ h^{-1})(168\ h))}$$

$c_3 = 0.07$ corresponding to a weight percent of 7%, well below the saturation concentration of salt water of 26.889 wt %.[5]

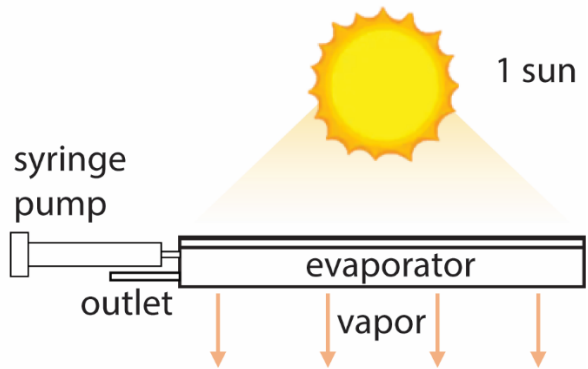

Figure S4. Evaporator-only test. The evaporator is pre-filled with salt water. A syringe pump reduces the flow rate until the flow rate of the discharged water at the outlet is negligibly low. Vapor exits the evaporator to the ambient environment.

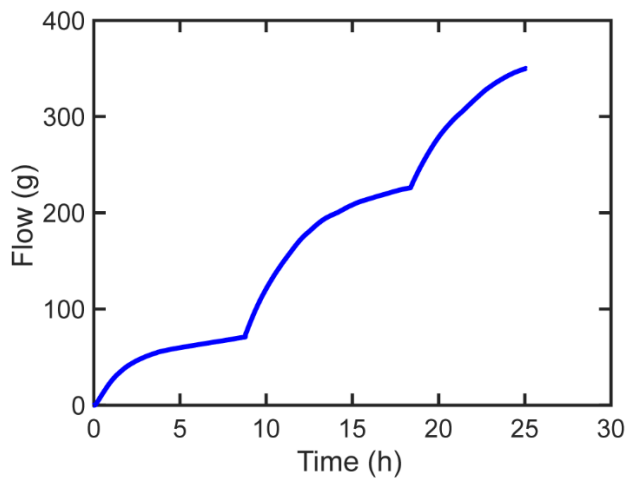

Figure S5. Flow out the gravity feeding bag during the 25 hour passive desalination test. Flow out of the gravity feeding bag was increased at hour 9 and hour 18, indicated by the increasing flow rate at hour 9 and hour 18. In total, 350 g of sea water exited the gravity feeding bag.